# Pressure dependence of phonon modes across the tetragonal to collapsed tetragonal phase transition in CaFe$_2$As$_2$

R. Mittal[1,2], R. Heid[3], A. Bosak[4], T. R. Forrest[5], S. L. Chaplot[2], D. Lamago[3,6], D. Reznik[3], K.-P. Bohnen[3], Y. Su[1], N. Kumar[7], S. K. Dhar[7], A. Thamizhavel[7], Ch. Rüegg[5], M. Krisch[4], D. F. McMorrow[5], Th. Brueckel[1,8] and L. Pintschovius[3]

[1]*Juelich Centre for Neutron Science, IFF, Forschungszentrum Juelich, Outstation at FRM II, Lichtenbergstr. 1, D-85747 Garching, Germany*

[2]*Solid State Physics Division, Bhabha Atomic Research Centre, Trombay, Mumbai 400 085, India*

[3]*Forschungszentrum Karlsruhe, Institut für Festkörperphysik, P.O.B. 3640, D-76021 Karlsruhe, Germany*

[4]*European Synchrotron Radiation Facility, BP 220, F-38043 Grenoble Cedex, France*

[5]*London Centre for Nanotechnology and Department of Physics and Astronomy, University College London, London WC1E 6BT, United Kingdom*

[6]*Laboratoire Léon Brillouin, CEA-Saclay, F-91191 Gif sur Yvette Cedex, France*

[7]*Department of Condensed Matter Physics and Material Sciences, Tata Institute of Fundamental Research, Homi Bhabha Road, Colaba, Mumbai 400 005, India*

[8]*Institut fuer Festkoerperforschung, Forschungszentrum Juelich, D-52425 Juelich, Germany*

The pressure dependence of a large number of phonon modes in CaFe$_2$As$_2$ with energies covering the full range of the phonon spectrum has been studied using inelastic x-ray and neutron scattering. The observed phonon frequency changes are in general rather small despite the sizable changes of the lattice parameters at the phase transition. This indicates that the bonding properties are not profoundly altered by the phase transition. The transverse acoustic phonons propagating along the c-direction are an exception because they stiffen very significantly in response to the large contraction of the c-axis. The lattice parameters are found to change significantly as a function of pressure before, during and after the first-order phase transition. However, the frequencies change nearly uniformly with the change in the lattice parameters due to pressure, with no regard specifically to the first-order phase transition. Density functional theory describes the frequencies in both the zero pressure and in the collapsed phase in a satisfactory way if based on the respective crystal structures.

## I. Introduction

The discovery of superconductivity in iron arsenic compounds has triggered a large-scale research effort [1-13] to explore the physical properties of these materials and to understand the mechanism responsible for superconductivity in these materials. Similar to many of the high-Tc cuprates, the undoped compounds show [3] a structural phase transition from a high temperature tetragonal to a low temperature orthorhombic phase, followed by an antiferromagnetic spin-density-wave (SDW) transition. Electron or hole doping suppresses these phase transitions and induces superconductivity at lower temperatures. In contrast to the cuprates, these compounds are semimetals and therefore metallic even without doping. At present, it remains unclear whether the change of the electron concentration by doping is essential for achieving superconductivity or whether the suppression of the phase transition into a magnetically ordered state is the main effect. Indeed, it has been found that superconductivity can be induced without doping by applying pressure. For $CaFe_2As_2$, $T_c$ as high as 10 K has been found at a moderate pressure of 3.5 kbar [5], while for $SrFe_2As_2$ and $BaFe_2As_2$, superconductivity is achieved at about 28 K at 3.2 GPa and 4.5 GPa respectively [6].

Neutron powder diffraction measurements have shown [4] that $CaFe_2As_2$ undergoes a structural phase transition at P=3.5 kbar at T=50K. The new phase remains tetragonal but the lattice parameters change dramatically at the transition for which reason the new phase is generally called the collapsed phase: the structural transition causes a decrease of the c-axis lattice parameter by ~10% and an increase in the a-axis parameter by ~ 2%. This phase is not magnetically ordered and it was thought to be the primary candidate for the occurrence of superconductivity with $T_c$ ~ 10 K. However, recent magnetic susceptibility experiments carried out under purely hydrostatic pressure conditions using He as the pressure medium indicate that superconductivity appears [11] in $CaFe_2As_2$ only under somewhat non-hydrostatic pressure conditions. Nevertheless, it seems that chemical doping is not indispensable for achieving superconductivity.

Present-day density functional theory (DFT) is only partially successful [7,13] in predicting the structural, vibrational and magnetic properties of the iron-arsenides. The calculations predict the existence of several competing phases with nearly the same free energy but in the end they fail to predict the correct phases as a function of pressure and temperature. In a previous investigation we have shown [8] that DFT gives a satisfactory account of the vibrational properties of $CaFe_2As_2$ if the calculations are based on the experimentally determined structure. If, however, the calculations are based on the structure obtained from minimization of the free energy, there is very serious disagreement between experimental and calculated phonon frequencies. The disagreement can be traced to the fact that the optimized structure is close to that of the collapsed phase and not to that of the zero pressure phase. In the present investigation, we used pressure to produce the structure close to the optimized structure of DFT, and then studied lattice vibrations by x-ray and neutron inelastic scattering. The x-ray experiments were performed at room temperature. In this case, a fairly high pressure (about 17 kbar) is necessary to induce the phase transition into the collapsed phase. An even higher pressure (about 25 kbar) is necessary to make the c/a-ratio close to the value predicted by DFT for ambient pressure. The maximum pressure achieved in our experiments was even considerably higher, i.e. 48 kbar. This allowed us to study the pressure-induced frequency changes also in a pressure range where $CaFe_2As_2$ is expected to show the normal behavior of a solid, i.e. not influenced by a near-by phase transition.

The pressure cell used in the neutron measurements could be cooled to low temperatures. Therefore, a much lower pressure was sufficient to bring the sample into the collapsed phase. The structure finally reached at P = 6 kbar and T = 120 K was not much different from that at p = 25 kbar and room temperature in the x-ray measurements. We will show that the phonon frequencies measured by x-rays or neutrons in the collapsed phase correspond quite well to those predicted by DFT for the optimized structure indicating that the vibrational properties of $CaFe_2As_2$ are largely governed by the crystal structure.

**II. Experimental**

Plate-like single crystals of $CaFe_2As_2$, which crystallize in the $ThCr_2Si_2$ tetragonal structure (space group *I4/mmm*), were grown from a Sn flux [12]. The inelastic x-ray experiments were carried out using the ID28 beam line at ESRF, Grenoble (France). For our measurements we have chosen an energy resolution of 3 meV, corresponding to an incident photon energy of 17794 eV using the Si (999) set-up. The dimensions of the $CaFe_2As_2$ crystals were reduced to about 40-50 microns in diameter, and about 20 microns thick. The crystal was characterized before and after loading into a membrane diamond anvil cell. Neon was used as pressure transmitting medium in order to ensure hydrostatic conditions. The pressure was determined using the ruby fluorescence line. The FWHM of crystal rocking curve was about $0.2°$, which indicates good quality of the crystal, and found to remain almost the same during the entire experiment up to the highest pressure of 48 kbar.

For the inelastic neutron scattering measurements we have used the same sample as in our previous measurements [8] for the study of the phonon dispersion relation at room temperature. It had dimensions of 15 mm × 10 mm ×0.4 mm. The experiments were carried out on the 1T1 triple axis spectrometer at the Laboratoire Leon Brillouin, Saclay, which is equipped with vertically and horizontally focusing monochromators and analyzers resulting in high neutron intensity. For the high pressure measurements, the sample was mounted in a gas pressure cell with helium gas as the pressure transmitting medium. The cell was loaded with a pressure of p = 6 kbar and then slowly cooled down to 120 K.

The measurements at ESRF as well as at LLB were carried out in the a-c scattering plane. The ab-initio calculation of one-phonon structure factors was used to select the most appropriate Bragg points for the detection of particular phonons.

**III. Computational details**

Calculations in the framework of density-function theory were carried out using the mixed basis pseudopotential method [14]. We employed norm-conserving pseudopotentials and a plane-wave cutoff of 22 Ryd, augmented by local functions at the Ca and Fe sites. Brillouin zone summations were done with a Gaussian broadening of 0.2 eV and 40 k points in the irreducible part of the Brillouin zone (IBZ). In this study, we only used the generalized gradient approximation (GGA) as given by Perdew, Becke and Ernzerhof (PBE) [15]. The linear-response technique was employed for the phonon calculations [16]. Dynamical matrices were calculated on a simple tetragonal 4×4×2 grid (15 q points in the IBZ) from which phonon dispersions were obtained by standard Fourier interpolation [17,18]. We note that all the calculations for the present study were non-spin-polarized because we were interested in non-magnetic phases only.

To simulate the pressure effects, theoretical phonon dispersion curves were obtained for two sets of lattice constants for the tetragonal I4/mmm structure: (i) the fully optimized structure with a=3.988 Å. and c=10.608 Å, and (ii) a compressed structure with a=3.942 Å and c=10.33 Å. In both cases, the internal structural parameter was fully relaxed.

## IV. Results

The x-ray measurements could not be started at zero pressure, because the cell had to be loaded first with a certain gas pressure which depends on the final pressure to be reached in this experiment. However, the pressure actually chosen (P = 2.5 kbar) was low enough to allow an easy comparison with the zero pressure neutron data. Indeed, we found little difference between the data taken at zero pressure by neutrons and at P = 2.5 kbar taken by x-rays. The pressure was then increased in two steps to P = 16.5 kbar, i.e. a pressure slightly below that inducing the phase transition into the collapsed phase. The next pressure was chosen as P = 18.3 kbar, i.e. slightly above the phase transition. We conclude from our determination of the lattice constants that the phase transition happened indeed in this pressure interval as expected from ref. [4,9] (see Fig. 1). Thereafter, we increased the pressure in two steps up to the maximum pressure achievable with this set-up, i.e. p = 48 kbar. At each pressure, several energy scans were performed at Q= (2.5 0 0), (201) and (300). At 25 kbar and 48 kbar we have carried out additional scans at Q = (2.25 0 0) and (2.75 0 0). Typical scans taken at Q = (201) as a function of pressure are shown in Fig. 2. The scans were fitted to several peaks using a Lorentzian profile. The peak positions, normalized to that observed at the lowest pressure, are plotted in Fig. 3 versus the relative volume.

The neutron measurements suffered from a very large background generated by the walls of the pressure cell. The background increased with energy reflecting the phonon density of states of aluminium, i.e. the material of which the pressure cell was made. Moreover, as aluminium is a coherent scatterer, this background showed structure in momentum transfer Q. In the end, only scans searching for phonons of the TA100 and the TA001 phonon branches in the low energy region up to about 10 meV could be evaluated with confidence. A typical energy scan of that sort is depicted in Fig. 4. The bulk of the measurements were carried out at P = 6 kbar and T = 120 K. A few further scans were taken during the slow cooling procedure. The results of these scans are included in Fig. 5. The other neutron results are depicted in Fig. 6 together with the high pressure x-ray results.

## V. Discussion

Two conclusions can be directly drawn from inspection of Fig. 3: (i) with the exception of the transverse acoustic modes propagating along c (TA001) modes, the frequency changes with pressure are rather moderate in spite of the enormous structural changes, and (ii) the phase transition does not stand out in the diagram. The latter point tells us that the relative volume seems to be the most important parameter determining the frequency changes, and that the phase transition itself does not entail particular changes of the bonding properties. The relatively weak response of most phonon modes to the pressure-induced structural changes can be qualitatively understood by a near-cancellation of the stiffening due to shrinkage of the c-axis and the softening due to an expansion in the a-b plane. The TA001 frequencies are an exception, because they seem to be sensitive to the c-axis lattice parameter only (see Fig. 5, which shows a compilation of both neutron and x-ray data).

There are two widely used approaches to explain pressure-induced phonon energy shifts: one is based on empirical interatomic potentials and the other on DFT calculations. Interatomic potentials work in general not very well for metals for which reason they are not tried here. We now turn to DFT calculations. A part of them were available prior to our experiments. Additional calculations were carried out for the highly compressed structure observed at P = 48 kbar. From earlier studies [8] we knew that DFT has a problem with $CaFe_2As_2$ – and other iron arsenides as well – in that the structure with the lowest free energy, i.e. the optimized structure, is far away from the experimentally observed structure at zero pressure and ambient temperature. Moreover, the parameters of the optimized structure depend sensitively on the approximation used for the exchange and correlation potential. It was found that the optimized structure calculated within the general gradient approximation (GGA) is closer to experiment than that calculated within the Local Density Approximation (LDA). Still, the optimized structure within GGA does not correspond to the zero pressure phase but rather to the collapsed phase. In fact, the lattice constants of the optimized phase (a = 3.988 Å, c = 10.608 Å) are very close to those observed at P = 6 kbar and T = 120 K (a = 3.984 Å, c = 10.60 Å) or at P = 25 kbar and T = 300 K ( a = 3.96 Å, c = 10.55 Å). Therefore, our high pressure data lend themselves to a direct comparison with the frequencies calculated by DFT within GGA for the optimized structure. As is shown in Fig. 6, the calculated frequencies agree indeed quite well with the 25 kbar/300 K and the 6 kbar/120 K data. Clearly, the agreement is not perfect but nevertheless, the calculations capture the general trends when going from the ambient pressure phase to the collapsed phase. To facilitate the assessment, we have included the calculated results which are obtained by imposing the experimental structure at ambient pressure conditions in Fig. 6. We know from a previous investigation [8] that such an approach leads to a satisfactory description of the phonon frequencies observed at ambient conditions. We have further included some zero pressure data for a direct comparison. Inspection of Fig. 6 shows that the exceptionally strong stiffening of the transverse acoustic modes in the 100- and the 001-directions is quantitatively reproduced by the calculations.

For pressures higher than 25 kbar, i.e. far above the phase transition into the collapsed phase [9], the c/a-ratio changes rather little. As a consequence, $CaFe_2As_2$ is expected to show the normal behavior of slightly anharmonic solids for such high pressures. Indeed, DFT predicts mode Grüeneisen parameters $\gamma = -(\Delta\nu/\nu)/(\Delta V/V)$ for this pressure range which lie mostly between 1 and 2. An experimental determination of mode Grüeneisen parameters is relatively difficult because the frequency shifts between 25 kbar and 48 kbar are generally quite small. The few $\gamma$ values which could be evaluated with confidence are in fair agreement with the predictions (Fig. 3(b)). In particular, the relatively large $\gamma$ value calculated for the zone boundary TA001 mode was confirmed by experiment. It reflects the fact that even for pressures above 25 kbar, the reduction of the lattice parameters is anisotropic, the c-axis being more compressed than the a-axis.

## VI. Conclusions

The structure of $CaFe_2As_2$ responds much more to pressure than most other metallic solids. In particular, the c/a-ratio changes enormously on applying a few tens of kbar. However, the vibrational properties of this compound change surprisingly little at the same time, except for a few vibrational modes which are very sensitive to the c-axis lattice parameter. It shows that the dynamical couplings are very similar for the different structures of $CaFe_2As_2$. The observed changes in the lattice dynamics as a function of the crystal structure are reasonably well described by DFT calculations although DFT does not predict the correct zero pressure structure. From this finding we conclude that this type of theory captures the bonding properties of this compound in a quite reliable way.

FIG. 1 (Color online) Temperature dependence of the a and the c axis lattice constants and unit cell volume as measured by x-ray and neutron diffraction. For comparison, the data published by Goldman et al Kreyssig et al. [4,9] are shown, too. The Kreyssig et al data were obtained on heating at a pressure of 6.3 kbar.

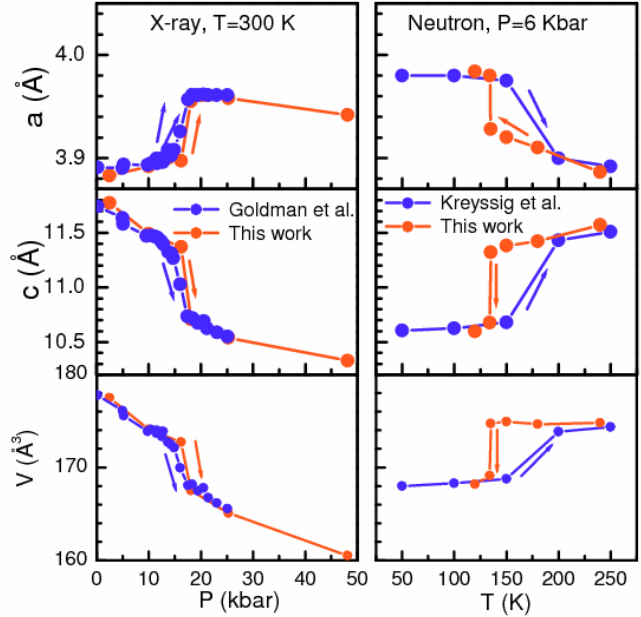

FIG. 2 (Color online) Energy scans performed at Q = (2.5 0 0) in the inelastic x-ray scattering experiments. The open circles and full lines correspond to the experimental data and fit curves, respectively. The phase transition happened between 16.2 kbar and 18.1 kbar, leading to a rather large volume change ($\Delta V/V = -0.03$) in spite of the small increment in pressure.

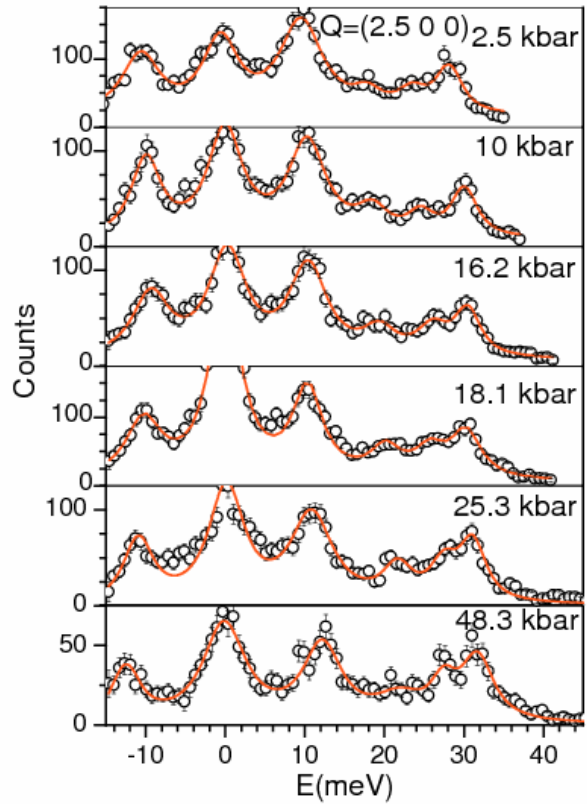

FIG. 3 (Color online) (a) The variation of phonon energies with volume. (b) Observed mode Grüeneisen parameters $\gamma_{obs}$ versus calculated mode Grüeneisen parameters $\gamma_{calc}$ in the pressure range above 25 kbar. $\Delta_{i,j}$ and $\Lambda_{i,j}$ represent the $j^{th}$ phonon mode in $\Delta_i$ and $\Lambda_i$ group theoretical representation respectively. The zone boundary modes of the TA001 and the first TO001 branch, labeled as $\Lambda_{3,1}$ and $\Lambda_{3,2}$ respectively, are at the same time the end points of the $\Delta_{2,1}$, $\Delta_{3,1}$ and the endpoint of the $\Delta_{1,2}$. The numbers in the bracket after the mode assignment gives the Q values for the measurements, while numbers after the Q values gives the phonon energies at ambient pressure. The dashed green line is drawn assuming a mode Grüeneisen parameter of 2 as expected for typical normal behaviour of solid. $V_C$ denotes the average of the volume before and after the transition from tetragonal to collapsed tetragonal phase of $CaFe_2As_2$.

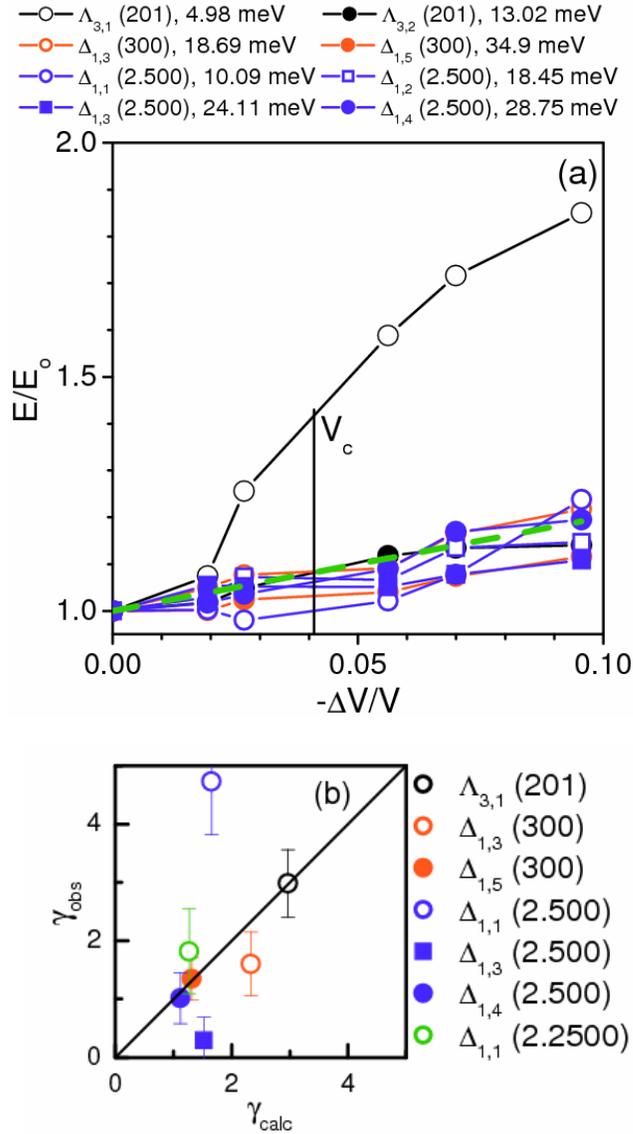

FIG. 4 (Color online) The energy scans obtained from the neutron inelastic experiments in the collapsed phase of $CaFe_2As_2$.

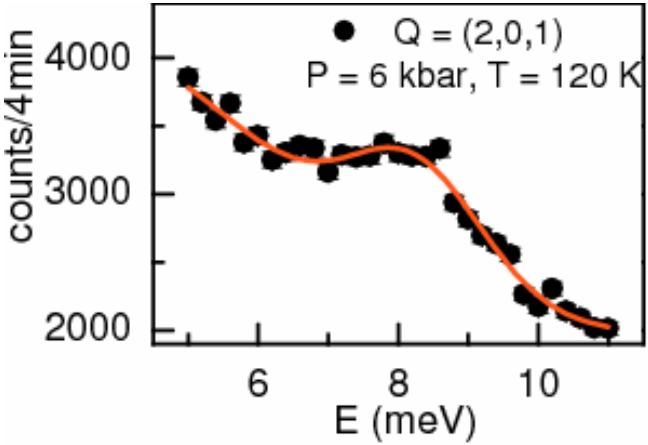

FIG. 5 Frequency of the zone boundary TA001 phonon versus c-axis lattice parameter as observed under various conditions in pressure and temperature. Full and open symbols refer to neutron and x-ray data, respectively. The broken line is a guide to the eye only.

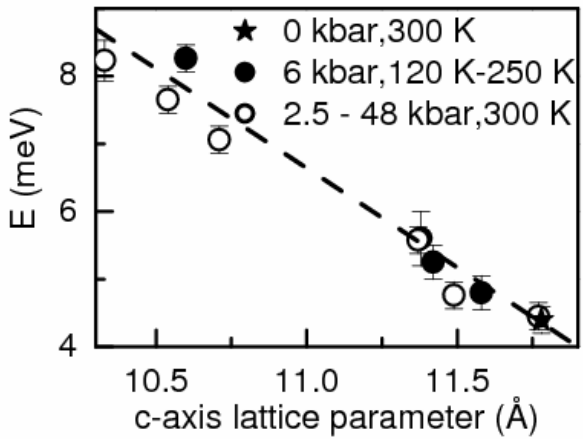

FIG. 6 (Color online) Comparison between the experimental and calculated phonon dispersion relation in ambient pressure and high pressure collapsed tetragonal phase of CaFe$_2$As$_2$. Results obtained from high pressure neutron inelastic measurements carried out at 6 kbar and 120 K are shown as well.

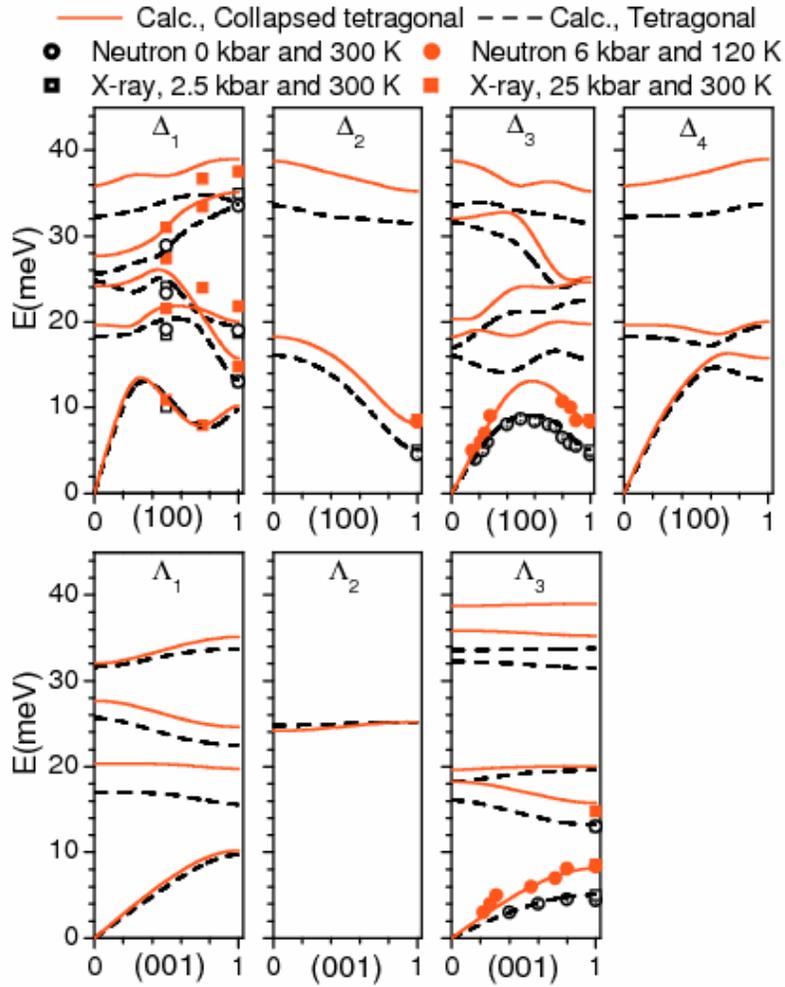